\documentclass[sigconf]{acmart}
\AtBeginDocument{%
  }

\copyrightyear{2026}
\acmYear{2026}
\setcopyright{cc}
\setcctype{by}
\acmConference[CHI '26]{Proceedings of the 2026 CHI Conference on Human Factors in Computing Systems}{April 13--17, 2026}{Barcelona, Spain}
\acmBooktitle{Proceedings of the 2026 CHI Conference on Human Factors in Computing Systems (CHI '26), April 13--17, 2026, Barcelona, Spain}
\acmDOI{10.1145/3772318.3791147}
\acmISBN{979-8-4007-2278-3/2026/04}

\acmBadgeR[https://www.acm.org/publications/policies/artifact-review-and-badging-current]{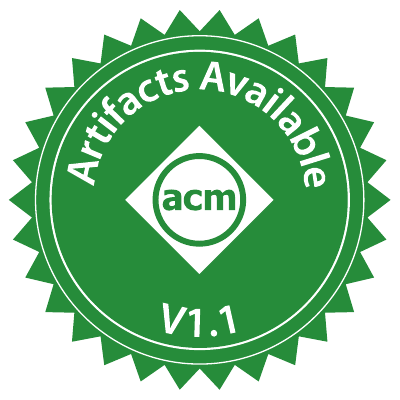}

\begin{document}

\title[Labor, Capital, and Machine]{Labor, Capital, and Machine: Toward a Labor Process Theory for HCI}

\author{Yigang Qin}
\orcid{0000-0001-7843-2266}
\affiliation{
  \institution{Syracuse University}
  \city{Syracuse, NY}
  \country{United States}
}
\email{yqin27@syr.edu}

\author{EunJeong Cheon}
\orcid{0000-0002-0515-6675}
\affiliation{
  \institution{Syracuse University}
  \city{Syracuse, NY}
  \country{United States}
}
\email{echeon@syr.edu}

\begin{abstract}

The HCI community has called for renewed attention to labor issues and the political economy of computing. Yet much work remains in engaging with labor theory to better understand modern work and workers. This article traces the development of Labor Process Theory (LPT)—from Karl Marx and Harry Braverman to Michael Burawoy and beyond—and introduces it as an essential yet underutilized resource for structural analysis of work under capitalism and the design of computing systems. We examine HCI literature on labor, investigating focal themes and conceptual, empirical, and design approaches. Drawing from LPT, we offer directions for HCI research and practice: distinguish labor from work, link work practice to value production, study up the management, analyze consent and legitimacy, move beyond the point of production, design alternative institutions, and unnaturalize bourgeois designs. These directions can deepen analyses of tech-mediated workplace regimes, inform critical and normative designs, and strengthen the field’s connection to broader political economic critique.

\end{abstract}

\ccsdesc[500]{Human-centered computing~HCI theory, concepts and models}

\keywords{labor process, work, management, exploitation, capitalism, Marxism, political economy}

\maketitle

\section{Introduction}

This article argues that HCI research and design should engage with labor issues as political economic questions—not only by documenting how people accomplish work with technologies, but also by analyzing and mitigating how design choices reshape labor conditions and relations in capitalist workplaces, conditions that affect many of us as researchers. To this end, we bring \textbf{Labor Process Theory (LPT)} to HCI as a structured framework for examining labor dynamics and technology under capitalism—one that helps problematize taken-for-granted design goals such as efficiency and productivity.

This reorientation challenges assumptions embedded in the field's history. Work and the workplace have historically been central to HCI/CSCW scholarship. From its roots in human factors research, HCI focused on optimizing how operators—typically workers—interact with machine interfaces to increase efficiency~\cite{kraft1982human, grudin2005three}. CSCW emerged as an interdisciplinary field aimed at understanding collaborative work practices and designing computing systems to support cooperation in workplace settings~\cite{schmidt2011concept, schmidt2000critical}. Rooted in this applied orientation, HCI/CSCW have largely viewed computing technologies as tools for practical outcomes: solving problems that hinder efficiency and productivity, or improving workplace experiences.

Designing such systems effectively required understanding the \textit{practice} of work: its concrete processes, actions, and interactions, particularly in CSCW. Several approaches emerged to meet this need. Activity Theory offered one lens~\cite{kuutti1991concept, wulf2018socio},  while a significant strand of early CSCW ``workplace studies,'' grounded in ethnomethodology and situated action, offered another~\cite{randall2021ethnography}. Scholars such as Lucy Suchman, Christian Heath, Paul Luff, and Dave Randall examined how people accomplish work through local, situated coordination---the micro interactional order. For these scholars, ``work'' referred to how actors produce social order through practice. This focus on work practice yielded designs better informed by workplace realities.

Yet this tradition of workplace studies was largely \textit{horizontal} in conceptualizing social relations: concerned with coordination, articulation work, and work organization among workers. What remained underexamined was the \textit{vertical} dimension—the structural relations of extraction and domination between labor and capital. These relations are rooted in the political economic setting of work: capitalism. As we elaborate in \S\ref{LPT review}, capitalism as a mode of production relies on extracting surplus value\footnote{In Marxist political economy, surplus value refers to the value created by workers in excess of their labor cost (wages), which is appropriated and reinvested by capitalists. Note that ``value’’ here is economic, distinct from the moral or cultural ``value’’ often discussed in HCI (e.g., Value Sensitive Design).} at the point of production, where workers, lacking means of production, must sell their labor power to capitalists. The gap between what they produce and what they receive as wages becomes the site of exploitation, producing a ``structured antagonism''~\cite{thompson2004labor} between labor and capital.

Critical HCI scholars have engaged with capitalism's broader implications, examining private property and commoning~\cite{chauhan2024commoning, lampinen2018member, beloturkina2025charting, sriraman2017worker, solano2025running, jack2025relational}, markets and supply chains~\cite{singh2025s, zhang2019designing, jack2025relational}, racial and colonial dimensions of capital~\cite{ogbonnaya2020critical, irani2010postcolonial, turner2023racial, mcmillan2020platform, kuo2022triangulating}, gendered divisions of labor~\cite{wajcman2000reflections, bardzell2010feminist}, socioeconomic class~\cite{ames2011understanding, yardi2012income, karki2025get, sambasivan2021re, tandon2025can}, ideology ~\cite{bardzell2013critical, lindtner2016reconstituting, lin2021techniques, so2025cruel, ahmed2022owns}, and ecological impacts of profit-driven development~\cite{sharma2025sustainability, sharma2023post, scuri2022hitting, nardi2017developing}. This body of work foregrounds structural inequality, power, and political economy. Yet such scholarship rarely examines how labor, capital, and machines are organized at the point of production—where value extraction, skill reconfiguration, and managerial control unfold through sociotechnical work systems.

This vertical dimension---attention to labor-capital relations in the workplace---was central to 1970s/80s Scandinavian trade union-based, participatory system design projects, which focused on workplace democracy and skill development~\cite{ehn1993scandinavian, kyng1988designing, greenbaum2020design}. Kjeld Schmidt, an influential scholar in European CSCW, was deeply informed by Marxism~\cite{schmidt2023conceptual}. Joan Greenbaum, an important political economist in the field, argued for a labor perspective: incorporating it would help researchers assess consequences of technological arrangements and understand how strategic industrial changes shape labor conditions~\cite{greenbaum1996back}. Over time, this perspective largely receded. Recent voices have called for revitalizing labor and political economic analysis~\cite{tang2023back, ekbia2015political, ekbia2016social, nardi2017developing}; emerging work has begun addressing labor issues directly~\cite{singh2023old, gloss2016designing, chen2022mixed, raval2016standing}. However, structural and theoretical labor analysis has yet to be systematically integrated into HCI research. 

Labor Process Theory (LPT) provides such a structural framework for HCI. Deriving from Marx's analysis in \textit{Capital} Volume I, LPT theorizes how labor power is transformed into labor in capitalist production processes~\cite{thompson1989nature}. LPT is particularly relevant to HCI, which studies and designs workplace technologies for production and coordination. Such technologies profoundly impact the working class and broader politics. LPT can explain these impacts. Its emphasis on the \textit{vertical} dimension—labor-capital relations—can help HCI attend to not only how work is coordinated but also how exploitation and domination are secured. It shows how control over the labor process can transform class conflict into competition among workers consenting to their own subordination. 
Through LPT, we can account for key mechanisms structuring the capitalist labor process: the extraction of surplus value, the degradation of work, the conception-execution separation, the deployment of managerial apparatus, and ideology and culture's effects on securing worker consent. LPT also addresses how these mechanisms have evolved in the post-industrial, digitized, and globalized world.

This article makes two contributions. First, we bring the labor process into focus as a critical \textit{site of analysis} for HCI—the site where capital organizes labor, technology, and management to secure surplus-value extraction. We trace LPT's development from Marx to contemporary scholars, making this theory accessible to HCI researchers.
Far from competing with existing critical lenses, LPT deepens theoretical foundations for analyzing technology, labor, and capital by linking critique of capitalism with the concrete organization of work---how tasks are divided, technologically mediated, and controlled in capital's interest. LPT complements existing HCI accounts of worker experience by situating them within the managerial and organizational mechanisms that organize exploitation.
Second, we review HCI literature through LPT's lens and propose directions for research and practice, outlining an agenda that directs attention to mechanisms the field has often sidelined: how work is commodified, how surplus value is extracted and obscured, how control is legitimized, and how supply chains and capitals become mobile---all mediated by technical assemblages.

This effort is critical as policy discourses increasingly promote a ``future of work'' vision that portrays AI and automation as inevitable and benevolent~\cite{riek2025future}. Proponents rarely question the labor-extraction regimes behind high-tech interfaces, the control apparatus that subjects workers to arbitrary discipline~\cite{alkhatib2019street}, or the employment precarity driven by cost reduction~\cite{alkhatib2017examining}. LPT, with its focus on labor-capital relations in production, equips HCI to address these issues---analyzing tech-mediated work regimes, informing critical and normative designs, and strengthening the field's connection to political economic critique.

The article proceeds as follows. We first introduce LPT by tracing its conceptual development as a foundational analysis of work under capitalism (\S\ref{LPT review}), with a particular focus on LPT's insights into machines (\S\ref{MachineIssue}). We then review HCI scholarship through the lens of LPT to map how the field has studied labor and technology (\S\ref{Findings}). Finally, we offer directions for HCI research and practice (\S\ref{Discussion}).

\section{Labor Process Theory: An Evolving Program} \label{LPT review}

LPT emerged in the 1970s when Harry Braverman revived and extended Marx's analysis of labor, capital, and wage work. Subsequent scholars contested Braverman's theses, producing successive waves of theory as work itself continued to transform. This section outlines key ideas marking LPT's development across waves, laying the foundations for integrating LPT with HCI. We begin with Marx, who gave the term \textit{labor process} its particular theoretical meaning.

\subsection{The ``Hidden Abode of Production'': Where Is Exploitation?}
\label{S2.1: hiddenAbode}

In \textit{Capital} Volume I, Marx argued that to understand exploitation, one must leave the market—that sphere of ``Freedom, Equality, Property and Bentham''—and enter the ``hidden abode of production''~\cite[pp. 279--280]{marx1990capital}. There, in the labor process, capitalism reveals its exploitative core.

Central to Marxian analysis is the distinction between \textit{labor} and \textit{labor power} because they reveal the nature of exploitation. \textbf{Labor power} is potential—``the aggregate of those mental and physical capacities'' a worker sets in motion to produce something useful~\cite[p. 270]{marx1990capital}. \textbf{Labor} is actual exertion. The labor process is where this potential actualizes, where labor power produces commodities with dual nature: \textbf{use-value} (utility) and \textbf{value} (realized as exchange-value in the market). Workers do not sell labor; they sell labor power—their capacity to work for an agreed period~\cite{braverman1998labor}. Capitalists' goal is to extract as much actual labor as possible from this purchase.
This gap between labor and labor power is where capitalism's structural exploitation occurs and is specific to capitalism. To understand why, Marx contends that this mode of commodity production is historically unique. Unlike feudalism, industrial capitalism is a mode of production in which the proletariat must sell their labor power in exchange for wages from the bourgeoisie, who owns the means of production: land, resources, factories, and tools.

\begin{figure*}[htbp]
    \centering
    \includegraphics[width=1.2\columnwidth]{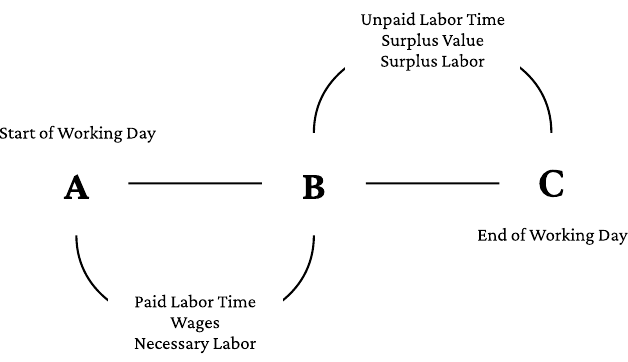}
    \caption{Value production and appropriation in the capitalist labor process. A and C represent the beginning and end of a working day, respectively. B is the hypothetical time point where the worker has produced value equal to their wage and begins producing surplus for the capitalist(s). In practice, B cannot be located on the timeline because wages are paid only when the working day ends.}
    \Description{This figure shows value production and appropriation in the capitalist labor process. Points A and C represent the beginning and end of a working day, respectively. Point B in between is the hypothetical time point where the worker has produced value equal to their wage and starts producing surplus for the capitalists. The A-B segment represents paid labor time, wages, and necessary labor; the B-C segment represents unpaid labor time, surplus value, and surplus labor. In practice, B cannot be located on the timeline because wages are paid only when the working day ends.}
    \label{valueProduction}
\end{figure*}

Capitalism's exploitation hinges on \textbf{surplus value}. Workers receive payment not for the total value they produce, but only for the cost of reproducing labor power, measured by the ``socially necessary labor time'' required to produce goods used for this reproduction~\cite{marx1978wagelabour}. 
Value beyond what workers' reproduction requires constitutes surplus value, appropriated by capitalists. Driven by competitive pressure for expansion, capitalists reserve part of produced value for reinvestment while paying wages at the minimal level to reduce costs. \autoref{valueProduction} illustrates this division of the working day—between necessary labor (for subsistence) and surplus labor (for capital). This exploitation is concealed because the wage contract appears as fair exchange, obscuring surplus value appropriation. Workers are exploited not because of low pay or poor conditions per se, but because they must generate surplus labor for capitalists to secure their subsistence.

How do capitalists maximize surplus value? There are two strategies. \textit{Absolute surplus value} is extracted by prolonging surplus labor time either by lengthening the working day (extensive exploitation) or increasing production pace and intensity (intensive exploitation). In \autoref{valueProduction}, lengthening the working day extends the A-C segment, thereby extending the surplus labor segment (B-C). \textit{Relative surplus value} is extracted by shortening necessary labor time (the A-B segment), thus extending the B-C segment. This occurs at the societal level when machinery and labor divisions increase average productivity, reducing the socially necessary labor time to produce subsistence goods and thus reducing labor power's value (wages).
These techniques yield profound consequences for the working class. Pursuing absolute surplus value expands work hours, while extracting relative surplus value systematically devalues labor power. This productivity push renders workers superfluous, causing layoffs and expanding the ``industrial reserve army''—the unemployed and underemployed population who, competing for scarce jobs, exert downward pressure on wages and bargaining power for all workers~\cite[p. 781]{marx1990capital}. 

This has implications for HCI: the very tools of efficiency, namely cooperation, division of labor, and machinery, can become instruments of intensified extraction; the common design goal of productivity gains can, at the societal level, devalue labor power even as it temporarily increases output.

\subsection{Deskilling, Scientific Management, and Control Imperative}
\label{S2.2: deskilling}

After Marx, analysis of the labor process lay dormant for decades until Harry Braverman revived it. Braverman extended Marx, showing how capitalists reorganize production to maximize surplus value through labor process control~\cite{braverman1998labor}. His critique of industrial sociology's focus on morale and productivity established LPT as a research program~\cite{littler1990labour, eldridge1991indsutrial, thompson2004labor}.

\textbf{Deskilling} is central to Braverman. Complex processes are broken into discrete, repetitive operations. Workers, responsible only for narrow steps, lose control over the whole production. This fragmentation reduces training needs, lowers labor costs, and cuts workers off from the overall process. Above all, it deepens alienation.

Equally critical is the \textbf{separation of conception from execution}. Managers monopolize planning, design, and coordination while workers are left with short-sighted execution. This division stripped workers of discretion, rendering their work less skilled and valued. Braverman contrasted this with traditional craftwork, where conception and execution were unified and workers retained autonomy.\footnote{Romanticizing craftwork is one criticism of the deskilling thesis~\cite{cutler1978romance}. See others~\cite{friedman1977industry, thompson1989nature, wood1982degradation}.} Under capitalism, this unity dissolved, reinforcing managerial authority and weakening worker agency.

\textbf{Taylorism}, or scientific management, systematized these principles. Through time-and-motion studies, Frederick W. Taylor decomposed complex tasks into standardized, measurable steps and prescribed optimal methods. This meant dividing tasks into finer, repetitive operations: shifting from a handful of workers performing multiple tasks in a cottage to hundreds executing single steps on a shop floor. It demands intensive supervision. Taylor dictated not only what workers did but how, eliminating ``superfluous and awkward'' movements, stripping discretion, expanding oversight. Taylorism thus exemplified how capitalist rationalization, under efficiency's guise, eroded skills while reinforcing control~\cite{taylor1919principles}.

The \textbf{control imperative} stems from labor power's peculiar nature: malleable yet embodied in humans whose will is never fully subordinated~\cite{friedman1990managerial}. This creates persistent uncertainty, compelling constant supervision to ensure that labor power transforms into labor. The primary mechanism, Braverman argued, was separating conception from execution: management monopolized planning while workers were confined to execution. As Friedman and Edwards observed, this imperative arises not from abstract desires for domination over workers, but from profit pursuit itself. The root problem is the gap between what capitalists buy (labor power) and what production requires (labor). Bridging this gap requires continual managerial control to convert labor power into labor effort~\cite{edwards1979contested}. Control thus becomes imperative.

Braverman established LPT's core mission: studying the antagonistic labor-capital relation within production. By showing how deskilling, conception-execution separation, and Taylorism serve surplus value extraction, he repositioned the labor process as central to understanding capitalist exploitation. Yet critics argued his framing cast workers as passive subjects, underplaying resistance and autonomy~\cite{thompson1989nature, wood1982degradation}. Post-Braverman theorists in the late 1970s and 1980s addressed this gap and complicated the labor-capital relation, producing the ``second wave'' labor process analysis~\cite{thompson2004labor}. 

For HCI, Braverman's insights suggest potential design directions: to name a few, systems that upskill or reskill workers, reunite conception and execution, and restore agency. They also cast doubt on purported universal benefits of divisions of labor coordinated by computers.

\subsection{Coercion, Consent, and Factory Regime}
\label{S2.3: control}


In response to Braverman's deterministic framing, which left little room for worker agency or everyday resistance, labor process analysts sought to remedy this absence. Drawing on case studies, they developed models of worker-management relations emphasizing diverse control modalities and worker responses. While emphasizing different facets, these accounts shared the aim of situating labor-capital antagonism in lived workplace dynamics. Together they formed the second-wave LPT~\cite{thompson2004labor}. Below we introduce key contributions by Andrew Friedman, Richard Edwards, and Michael Burawoy.

\paragraph{Andrew Friedman} 
Examining managerial strategies, Andrew Friedman argued that managers confront the problem of extracting surplus under conditions of worker resistance and organizational complexity. He distinguished two control strategies: \textbf{direct control}, relying on coercion, close supervision, and minimal worker discretion, and \textbf{responsible autonomy}, granting workers bounded independence to channel initiative toward firm goals. Responsible autonomy, Friedman wrote, ``attempts to harness the adaptability of labour power by giving workers leeway'' while direct control ``treats workers as though they were machines''~\cite[p. 78]{friedman1977industry}. Friedman emphasized that while Marx and Braverman analyzed direct control—especially Taylorism—worker resistance and coercion's limits led managers to rely increasingly on responsible autonomy to sustain surplus extraction.

\paragraph{Richard Edwards} 
In \textit{Contested Terrain}, Edwards advanced a control typology distinct from but complementary to Friedman's framework, though both addressed the balance of worker autonomy and managerial authority~\cite{friedman1990managerial}. Control system, Edwards argued, involves direction, evaluation, and discipline. He then traced how these elements constitute three dominant forms of control: simple, technical, and bureaucratic.

\textbf{Simple control} prevails in small firms, where entrepreneurs and foremen exercise personal authority through threats, favoritism, hiring, and firing. Edwards described this as rule by despotic supervisors relying on bullying and arbitrary sanctions~\cite{edwards1979contested}. As firms expand, personalized authority extends through hierarchies, yet different forms also emerge. \textbf{Technical control} starts with machinery and scientific planning. Edwards noted that technical control works by ``designing machinery and planning the flow of work to minimize the problem of transforming labor power into labor'' and to maximize efficiencies~\cite[p. 112]{edwards1979contested}. Chaplin's \textit{Modern Times} captures it vividly: workers become machine attendants, discretion stripped by the assembly line's relentless pace. Technical control's success in managing blue-collar work transfers to clerical labor through information systems, such as keystroke tracking and screen monitoring~\cite{thompson1989nature}.

As firms expand, \textbf{bureaucratic control} emerges as the dominant form, institutionalizing authority through formal rules, classifications, and career ladders, enabling management to direct, evaluate, and discipline workforces systematically and impersonally. ``Rule of law—the firm's law—replaces rule by supervisor command,'' as company policy governs evaluation, promotion, and sanctions~\cite[p. 21]{edwards1979contested}. This system promises workers a ``career,'' but bounds them tightly to organizational hierarchies. Edwards's framework illustrates how control became progressively formalized and rationalized, embedding managerial authority in capitalism's structures.

\paragraph{Michael Burawoy}

Like other second-wave analysts~\cite{cressey1980voting}, Burawoy rejected the despotic, deterministic characterization of capitalist control. Drawing on Gramsci's concept of \textit{hegemony}, he argued that domination operates not only through \textit{coercion}—overt authority workers experience as hostile—but also through \textit{consent}, where management ``wins over'' workers and co-opts their initiative with corporate ends~\cite{ramos1982concepts}. Burawoy applied this distinction between coercion and consent to examine a machine shop floor.

In \textit{Manufacturing Consent}, Burawoy showed how workers were drawn into the game of ``\textit{making out}''—meeting piece-rate quotas for bonuses, recognition, and ``relative satisfactions'' of playing the game itself~\cite{burawoy1979manufacturing}. The labor process becomes a site of \textit{consented} exploitation, as workers found genuine satisfaction in beating quotas and outpacing peers. Workers appeared to act autonomously, yet their participation---including forms of resistance---reproduced \textbf{consent} to surplus value extraction.
To further his distinction on despotism and hegemony, Burawoy in \textit{The Politics of Production} identified types of \textbf{factory regime}—bureaucratic despotism, market despotism, and hegemonic regime—and the political conditions of state and market enabling each~\cite{burawoy1990politics}.

Burawoy extended labor process analysis by showing how control operates through the subjective as well as the objective, harnessing workers' pursuit of autonomy. His argument inspired wide discussion, though critics later questioned his framing of  worker subjectivity from postmodernist perspectives.\footnote{Knights~\cite{knights1990subjectivity} argued that Burawoy's treatment of subjectivity relied on humanist assumptions derived from early Marx, equating autonomy with psychological needs for creativity and dignity. This, Knights suggested, prevented fully theorizing worker subjectivity. Others criticized Burawoy's neglect of the effects of gender ideologies on worker subjectivity and consent (e.g.,~\cite{cockburn1983brothers, cockburn1985machinery, collins2000gender,acker2006class}).} 

These three analysts shaped LPT's second wave as a theory of control, resistance, accommodation, and consent within factory regimes~\cite{edwards1986conflict}. LPT-inspired work in the '90s examined new management practices: post-Fordism promising workplace flexibility, teamwork and conceptual tasks~\cite{vallas1999rethinking}, and lean production (Toyotaism) rationalizing cost calculation and resource allocation to fine-grained levels~\cite{graham1995line}. Despite these changes, third wave LPT affirmed rather than revised the second wave's framework~\cite{thompson2004labor}. The second and third waves accumulated case studies, decomposing managerial strategies, control mechanisms, and worker resistance. This period also saw exploration of gendered and racialized dimensions of the labor process, leading to intersectional analyses~\cite{cockburn1983brothers, cockburn1985machinery, collins2000gender, acker2006class}. 

For HCI, the second-wave concepts of consent, control, and resistance illuminate contexts such as platform work and digital piecework. In gig works, quotas and gamified features like badges, titles, rankings, and rewards provide a sense of achievement as an incentive and promise autonomy while drawing workers toward ``voluntary,'' consented overwork. These concepts also remind that a regime of control combine multiple means---technical, bureaucratic, and normative---rather than relying on any single mechanism such as algorithms alone.

\subsection{Post-Industrialization, Platformization, and Global Production}
\label{S2.4}

The fourth wave addressed labor in service work, knowledge work, platform work, and global production networks. This demonstrated the continued relevance of core LPT concepts, such as control, value extraction, and division of labor, for understanding emergent work arrangements.

Service work collapses the distinction between production and consumption---the service is consumed as it is produced. A call center agent's labor, for instance, exists only in the moment of interaction. Because demand is volatile, temporality becomes contested terrain for profitability~\cite{negrey2025temporal}. 

Beyond temporal pressures, service work also requires management of emotions. Sociologists have highlighted emotional labor—``the management of feelings to create a publicly observable facial and bodily display'' that is ``sold for a wage''~\cite[p. 7]{hochschild2012managed}—and aesthetic labor, appearance management for meeting organizational demands~\cite{mears2014aesthetic, williams2010looking}. Yet as the concept of emotional labor has been stretched beyond its analytical boundaries~\cite{bolton2010old}, LPT re-centers these within political economy, distinguishing emotional labor that produces value for capital~\cite{bolton2010old} from emotion work present in everyday social interactions~\cite{hochschild1979emotion, goffman2023presentation}. Moreover, service workers' smiles are both emotional and bodily performances, linking to debates on embodiment in the labor process—in both productive spheres (manufacturing, knowledge work) and reproductive ones (domestic work, healthcare)~\cite{knights2017embodying, wolkowitz2017embodying}.

Meanwhile, the knowledge economy promised greater autonomy~\cite{thompson2024labour}. But in reality, the separation of conception from execution persists.
Physicians' execution is outsourced to nurses, lawyers' to paralegals, and professors' to student assistants~\cite{negrey2025temporal}. AI increasingly substitutes for much of the execution labor. Electronics industries exemplify a ``Bravermanian product cycle''~\cite[p. 193]{mcintyre2017marxian}: design is concentrated in core firms while assembly is outsourced to low-wage manufacturing sites.  Such trends appear among game developers with limited autonomy under supervisory procedures~\cite{Deuze2007professional}, tech workers under ``digital Taylorism''~\cite{liu2023digital}, and pharmaceutical workers whose tacit knowledge is codified and appropriated by management under financial pressures~\cite{mackinlay2002limits}.

Alongside knowledge work, platform work emerged as a major site of fourth-wave analysis. Studies proliferated across ride hailing, food delivery, crowd work, and freelancing. LPT concepts proved effective in analyzing these arrangements~\cite{chen2024digitalrun}: algorithmic management as normative control~\cite{kellogg2020algorithms}, micro work as new exploitable occupations, and informal resistance as a persistent response~\cite{thompson2024labour}.

Beyond individual workplaces, labor process analysts have extended their focus to global production networks, where corporations distribute production across national borders in dispersed yet interconnected sites~\cite{thompson2024labour}. In these networks, logistics~\cite{moore2018paying} and labor mobility~\cite{smith2010go} have beome key areas for examining how value and labor power flow across borders. For transnational value chains, analysts have incorporated governance and state coordination to explain why labor processes vary across national contexts~\cite{taylor2017globalization}, while also examining how financialization intensifies surplus value extraction, as pressures to maximize shareholder value tighten workplace performance demands globally~\cite{thompson2013financialization, cushen2016financialization}. 

These studies direct HCI researchers' attention to contexts such as logistics, transnational supply chains, financialized workplaces---where technologies mediate value extraction across borders, yet remain underexamined in computing research.


\subsection{Summary}
Despite new employment forms and production sites examined in \S\ref{S2.4}, LPT's core claims remain relevant for today's world economy. Here we summarize LPT's central arguments. We draw on the ``core theory'' from the 1990s, when LPT theorists distilled four principles~\cite{thompson2004labor, thompson2010capitalist}.\footnote{One recognized limitation is LPT's focus on labor transformation at the expense of  the social reproduction of labor power. Feminist scholars have offered critiques~\cite{gibson1997end, mezzadri2023social}. Yet as unpaid reproductive labor becomes paid productive labor—often performed by migrant women—LPT can be productively extended with a gender lens~\cite{wolkowitz2017embodying}.} 

\begin{itemize}
    \item The labor process is where \textbf{surplus value} is produced, and thus holds a privileged position in political economy (\S\ref{S2.1: hiddenAbode});
    \item Market competition impels capitalists to constantly revolutionize the \textbf{forces and organization of production} (\S\ref{S2.2: deskilling});
    \item Managerial control is necessary for capitalists to reduce \textbf{indeterminacy} in transforming labor power into labor (\S\ref{S2.2: deskilling});
    \item Labor and capital are in a ``structured antagonism''~\cite{thompson2004labor}. Yet capital, to revolutionize the labor process, must also seek initiative and cooperation from workers. This tension results in a continuum of worker responses—from \textbf{resistance} to accommodation and \textbf{consent} (\S\ref{S2.3: control}).
\end{itemize}


\section{The Machine Issues in Labor Process Theory} \label{MachineIssue}

Technology has always been central to LPT—inseparable from questions of value extraction and labor-capital relations~\cite{hall2010renewing}. This section highlights how LPT theorizes machines, with contemporary examples demonstrating relevance to HCI.

For Marx, machinery is a ``\textbf{labor-saving}'' technique~\cite{knights1990introduction}. But labor-saving for whom? Machines compress the labor time necessary for production, creating market advantages for capitalists. But following the labor theory of value outlined previously—wherein the value of a commodity equals the socially necessary labor time required for its production—each product's value decreases as more productive technologies become pervasive. This product devaluation drives down workers' necessary consumption costs for reproducing labor power, ultimately reducing their wages. Moreover, machines save labor per commodity, not per day. As productivity rises and profit rates fall, capitalists must sell more products—demanding more from workers, not less. The vision where AI works so that humans enjoy abundant leisure is a fantasy. More likely, humans will use AI to work in the future as much as they use computers today. ``Labor-saving'' machines rarely save labor.


Machinery also enables \textbf{deskilling}, reducing capital's dependence on specialized labor~\cite{thompson1990crawling}. The desire to control expertise incentivized capitalists to introduce machines. Marx traced this transition through two historical stages. During the \textit{manufacture} stage, capitalists relied on workers' tacit handicraft skills. Workers possessed embodied knowledge inaccessible to management, making them difficult to replace. Machinery changed this. Skills and knowledge were codified into machines—what Steinhoff calls ``absorption'': extracting qualities from labor and embedding them in fixed capital~\cite{steinhoff2024universality}.
During the \textit{machinery} stage, machines paced workers and defined tasks~\cite{friedman1977industry}. Production became ``the technological application of science,'' reducing labor to ``a mere moment of this process''~\cite[p. 281]{marx1978grundrisse}. Workers became interchangeable machine-tenders~\cite{resnikoff2022labor}. Unable to monopolize expertise, workers lost bargaining power~\cite{foley1986understanding} and working conditions eroded. Technological development, in this view, reflects class struggle~\cite{edwards1979contested}.

This logic—capital advancing technology to formalize and appropriate workers' knowledge—persists today. Contemporary examples include Knowledge Management Systems that extract, store, standardize, and distribute best practices and tacit knowledge~\cite{taskin2015knowledge} and machine learning pipelines that ``absorb'' skills and knowledge from data~\cite{steinhoff2024universality}. Yet the relationship between technology and deskilling is neither linear nor deterministic, as critics of Braverman's deskilling thesis have argued. New technology introduction can simultaneously create demands for new skills (reskilling) even as it fragments or downgrades existing work~\cite{wood1982degradation}. Effects vary across skills, occupations, experience, and modes of use. Contrary to predictions that automation uniformly reduces skill requirements, Holm and Lorenz, for instance, found that AI used as an input source increases learning in high-skilled jobs, while AI used as an order-giver reduces worker autonomy across skill levels~\cite{holm2022impact}.

Machines' deskilling effect is a key mechanism of \textbf{control}. When machinery transforms craft into repetitive tending, it eliminates worker discretion and dismisses their skill. Since deskilled workers are replaceable, workers' collective action becomes less potent. A twin effect---the separation of  conception from execution---requires layers of administrative technology that together form an architecture of control. This creates a situation where new ``labor-saving'' technologies necessitate more human and technological oversight to manage the increasingly complex, machine-driven workflow, thus requiring more extensive value extraction from workers. Thompson noted that control is rarely the stated purpose of new technology, but the desire to manage labor and extract value often serves as unstated justification for investment~\cite{thompson1990crawling}. 

Capitalists' drive to expand extraction's scale and depth motivates technological adoption. As Littler showed, new technologies address production ``bottlenecks'' such as delays, volatile demand, and coordination breakdowns~\cite{littler1990labour}. With each new technical solution to such problems, methods for controlling the expanded labor force were correspondingly transformed and intensified~\cite{littler1990labour}. Platforms and algorithms exemplify this dual movement today: absorbing large workforces while maintaining granular control over individuals. Behind the technological façade lies repackaged work intensification~\cite{hughes2025assembly}. This logic explains why Amazon weaponizes algorithms against worker organizing~\cite{wiggin2025weaponizing}. By emphasizing continuity of capitalist interest across technologies, LPT challenges tech-determinism and singularity narratives~\cite{joyce2023new}.

Finally, machinery intensifies workers' \textbf{alienation} from the production process, product, themselves, and social relations. The machine, as fixed capital and the product of others' labor,  confronts the worker not as a tool but as capital's physical embodiment. Knowledge and skill---once the workers' own---appear as alien forces. Living labor is subordinated to the machine's predetermined rhythm, ``subsumed under self-activating objectified labour''~\cite[p. 281]{marx1978grundrisse}. Workers' creative effort no longer directs production; it serves a mechanical system embodying past alienated labor and knowledge. Workers become superfluous except where capital requires them. 

Alienation has been a persistent interest in today's labor studies and has gained new dimensions through new technologies. For instance, Search Engine Optimization specialists feel powerless and insecure being held accountable for algorithmic performance indicators and Return on Investment (ROI) over which they have no control~\cite{ivanova2025alienation}. Platform work deepens workers' alienation from themselves and from social relations by constructing neoliberal, entrepreneurial subjectivities, manufacturing new types of consent that absorbs workers into fiercer competition, atomization, and exploitation~\cite{liang2025content, li2025re, galiere2020food, liu2025mercy, miszczynski2025coercion}.

These mechanisms, including exploitation, deskilling, conception-execution separation, and  regime of control, provide tools for investigating workplace technologies. We now turn to HCI scholarship to demonstrate the analytical value of these mechanisms.

\section{Labor Studies in HCI} \label{Findings}

To assess labor studies in HCI and highlight LPT's potential contributions, we conducted a semi-systematic literature review~\cite{snyder2019literature}. Semi-systematic review balances rigor with flexibility, suited for identifying themes and methods~\cite{snyder2019literature}. This suits our goal: surveying labor-related studies through LPT's lens to identify themes, approaches, and future directions. LPT serves as our guiding lens, structuring prior work to reveal how HCI has analyzed and intervened in labor-capital relations. Our review is illustrative rather than exhaustive.

\subsection{Search Strategy and Analysis}

We searched ACM CHI, CSCW, PACMHCI, TOCHI, DIS, CHIWork, PDC, ECSCW, and JCSCW for papers containing ``labor'' or ``labour'' in titles or abstracts.\footnote{The digital libraries of ECSCW and JCSCW do not support abstract filtering. Instead, we did full-text searches of the terms and manually screened the titles and abstracts for term use.} 
We focused on titles and abstracts because full-text searches returned overly broad results. 
The searches returned 483 papers across these venues. We manually screened abstracts, excluding papers that use ``labor'' nominally (e.g., ``labor-intensive'') and do not substantively engage with working conditions, labor relations, or wages. This filtering resulted in 223 relevant papers.

\begin{table*}[htbp]
\centering
\begin{tabular}{cccccccc}
    \toprule
    \textbf{Search Round} & \textbf{Terms} & \multicolumn{3}{c}{\textbf{Search Match}} & \multicolumn{3}{c}{\textbf{Filtered Match (Corpus)}} \\
    \cmidrule{3-8}
     & & ACM & E/JCSCW & Total & ACM & E/JCSCW & Total \\
    \midrule
    Primary Search & labor, labour & 452 & 31 & 483 & 207 & 16 & \textbf{223} \\
    \addlinespace
    \parbox[c]{0.08\textwidth}{\centering Additional Search} & \parbox[c]{0.3\textwidth}{\centering worker, manager, employee, employer, working class, workplace, capital*, Marx*} & \parbox[c]{0.06\textwidth}{\centering 914} & \parbox[c]{0.06\textwidth}{\centering 38 (title)} & \parbox[c]{0.06\textwidth}{\centering 952} & \parbox[c]{0.06\textwidth}{\centering 88} & \parbox[c]{0.06\textwidth}{\centering 2} & \parbox[c]{0.06\textwidth}{\centering \textbf{90}} \\
    \addlinespace
    \cmidrule{3-8}
    Total &  & 1366 & 69 & 1435 & 295 & 18 & \textbf{313} \\
    \bottomrule
\end{tabular}

\caption{Overview of the primary and additional search results. The * sign is a wildcard symbol which searches for words starting with the same preceding letters. The search terms are also exact match so that ``work'' is not returned when searching ``worker.''}
\label{Table:searchTerms}
\end{table*}

Recognizing that ``labor'' alone might miss relevant work, we conducted an additional search. We searched in above venues for titles and abstracts that contain a list of relevant terms but not ``labor'' or ``labour'' to collect new relevant papers. We curated search terms by consulting LPT-oriented sociology papers and testing iteratively. We eventually included core actors (worker, manager, employer, employee) and space (workplace) while avoiding polysemous terms like ``work'' and ``employment'' that generate false positives. We also included ``capital*'' and ``Marx*'' to capture papers engaging capital explicitly. The search terms and overview of the results are shown in \autoref{Table:searchTerms}. The criterion for relevance was the same as that of the primary search. This added 90 papers, expanding the corpus to 313 (listed in Supplementary Material). As expected, the additional search returned articles less focused on labor issues but complementary to the primary search. Our searches cover publications through November 2025. This means work after this time is not included. 

The first author read selected papers, maintaining notes and analytical memos. Emerging interpretations were discussed iteratively with the second author. Rather than line-by-line coding, the first author selectively read labor-relevant sections, open-coding and categorizing by occupation (e.g., data work), labor issue (e.g., invisible labor), and methodology (e.g., describing work challenges). Papers were compared within and across categories to reveal similarities, connections, and differences.

Two questions structured the analysis: \textit{What labor themes have HCI scholars examined?} and \textit{What research approaches have they used?} We used LPT as the guiding framework to appraise HCI work, reflecting on the gaps and opportunities in current research. When reviewing themes, LPT sensitized our attention to how structured antagonism gets into pairs of dichotomies in HCI literature: control versus resistance, individual tactics versus collective action, social versus technical control, and labor versus work. Tensions between LPT and HCI's orientations arose when reviewing approaches, but these turned out to be generative. LPT as a structural, explanatory theory has distinct purposes, qualities, and functions from those of HCI. The first author initially struggled to align the two field's orientations. But contrasting LPT with HCI ultimately clarified HCI's capacities and limitations, guiding us toward adapting LPT for HCI's purposes. Specifically, LPT provides political economic grounding for studying invisible work. It also offers explanatory power that complements HCI's descriptive norms and interventionist orientation. The narrative that follows reflects these focal points and tensions.\footnote{We did not cite all papers reviewed. Selections reflect thematic representativeness and saturation, privileging papers that best illustrate the themes.}

\subsection{Main Themes of Labor Studies}


\subsubsection{Control and Management}

Managerial decisions determine what tasks workers perform, how, and at what pace. HCI research frames this directive function as a key control mechanism. For example, \citet{fox2019managerial} found that managerial rationale around labor optimization and cost reduction drove IoT system introduction in public restrooms. In gig work, algorithms assign tasks and structure workflows~\cite{anjali2021watched}, though pace and intensity vary across platforms and contexts~\cite{sekharan2025designing, kusk2022working}. \citet{jarrahi2020platformic} extend this by revealing that digital labor platforms control workers to maintain transaction viability and protect platforms from disintermediation between freelancers and clients.

Monitoring is a well-established control technique in the labor process. HCI research has analyzed how digital systems monitor, evaluate, and affect workers. Monitoring operates across multiple channels, forming a technological assemblage that tracks workers~\cite{bakewell2018everything}. Contemporary monitoring is also more passive, mediated by ``stochastic machine witnesses''~\cite{gould2024stochastic}. Beyond passive monitoring, digital systems in blue-collar workplaces increasingly require workers not only to perform tasks but to document them. This traceability, designed to attribute failures to specific steps, shifts accountability onto workers~\cite{kristiansen2018accountability}. Management also employs peer-to-peer surveillance---sometimes formalized through systems like Total Quality Management~\cite{klein1993social}---encouraging lateral monitoring and relational control among workers~\cite{bakewell2018everything, friedman1990managerial}. The psychological toll of surveillance—fear, stress—also appears in HCI scholarship~\cite{doggett2024migrant}. Such data-driven supervision often culminates in disciplinary measures against workers, such as termination~\cite{singh2023old}.

Beyond direct supervision, HCI studies have examined subtler, ideological forms of control. Drawing on LPT, \citet{singh2023old} show how platform labor control combines algorithmic management with ``neoliberal social control.'' Neoliberalism constructs workers as self-responsible individuals whose compliance is secured not by explicit orders but by internalized promises of flexibility and autonomy. This underscores the need to analyze technical infrastructures alongside governmentality. Other studies trace how new technological environments reactivate familiar discipline. \citet{alkhatib2017examining}, for instance, show that food delivery work replicates piecework-like dynamics, where earnings tie to output-based quotas—recalling past shop floor regimes. \citet{sehrawat2021everyday} note that platforms like Uber manufacture compliance and stability through routinizing drivers' everyday app interactions. \citet{kim2025decoding} further illustrate how Upwork freelancers’ particular meanings of career success are shaped by platform logics of visibility, ranking, and evaluation.

HCI has also examined new domains and actors of control. \citet{lu2021uncovering} examined employers using social media to hunt for committed, job-ready low-wage workers, reproducing power relations in hiring. Client reviews, standardized skill sets, and worker status classification are recognized as new control forms~\cite{holtz2022much, munoz2022platform, anjali2024entangled}. Besides platforms, new actors such as guilds~\cite{yang2024guilds} and AI developers~\cite{sambasivan2022deskilling} can be sources of control over data workers. 

Much work here could benefit from LPT concepts like responsible autonomy and consent that explain how workers come to conform to or resist the discipline in the face of social and ideological means of control. Given control's complexity, opportunities exist to examine control across physical, digital, and social arenas, as illustrated by LPT's regime of control (see \S\ref{Dis 5.3: management}).

\subsubsection{Worker Resistance and Collective Action}

HCI scholarship has documented workers' micro-level resistance to control. One common form is refusing workplace technologies. Workers avoided company channels and used digital tools to circumvent surveillance~\cite{bakewell2018everything, saether2023workers}. Similarly, London bus drivers chose not to activate performance monitoring systems installed on buses~\cite{pritchard2015drive}.

Beyond refusal, workers deploy subtler tactics within systems they cannot avoid. London bus drivers engaged in strategic underperformance, driving just well enough to keep expectations low~\cite{pritchard2015drive}. MTurk workers reduced rejection risk by avoiding bad clients, sharing experience, and selectively taking tasks~\cite{mcinnis2016taking}. \citet{dombrowski2017low} identified precarious workers' sociotechnical strategies to address wage theft, including identifying wage discrepancies, documenting work, and pursuing wage claims. \citet{cheon2024examining} documented Amazon workers' reactions to algorithmic metrics, arguing reactivity functions as agency.

Everyday resistance expresses complaint but may be limited as incidental, ephemeral efforts seeking cracks of comfort, unable to subvert dominant power relations~\cite{edwards1990understanding}. Moreover, by gaining minimal satisfaction through management's loopholes, workers may pursue individual advantages over solidarity~\cite{burawoy1979manufacturing}.

Besides individual resistance, HCI researchers have explored technology for collective action. For instance, \citet{hsieh2024worker} discussed ways of building worker data collectives to strengthen collective voice; \citet{qadri2021s} appraised the potential of informal mutual aid for building gig worker solidarity; \citet{so2025cruel} suggested mobilizing tech workers' affective attachment to industry for collective resistance to layoffs. \citet{khovanskaya2019tools} proposed securing wage transparency, data contestation, and selective participation in management decision-making as potential tactics for protecting digital labor. Researchers have also sought to explain collective action's successes and failures. \citet{salehi2015we} found MTurk worker discussion on a platform they designed alternated between stalling and friction, impeding collective action. By contrast, \citet{thuppilikkat2024union} studied digital labor unionism in India and found the success factors to be utilizing technology, centering worker-organizers, and resisting platforms and the state. 

From an LPT perspective, the labor process, including technology arrangements, results from ongoing workplace struggles. Beyond how technology facilitates collective action, HCI researchers could study how mobilization effects reconfigure technology use subsequently, and how and why different mobilizations place different weights on technologies in the hope that such knowledge can inform design that supports organizing efforts strategically.

\subsubsection{Emotional and Body Labor}

Emotional and body labor are now widespread concepts across disciplines~\cite{bolton2010old} including HCI.
HCI empirical studies explored emotional labor of violence consultants~\cite{tseng2021digital}, home care workers~\cite{poon2021computer}, Uber drivers~\cite{gloss2016designing, kameswaran2018support, raval2016standing}, content moderators~\cite{dosono2019moderation, steiger2021psychological, wohn2019volunteer}, academic researchers~\cite{wolters2017emotional, balaam2019emotion}, Airbnb hosts~\cite{nemer2018airbnb}, Wikipedia editors~\cite{menking2015heart}, platform-based gig tutors~\cite{zhang2025identity}, and AI-monitored employees~\cite{roemmich2023emotion, corvite2023data}. Much emotional labor studied in the service sector is performed by women or \textit{feminized roles}~\cite{huber2022approximation}.
In response, design efforts have developed assistants and intervention systems providing emotional support for front-line workers such as home caregivers~\cite{bartle2022second}, front-office personnel~\cite{das2025ai}, violence consultants~\cite{chau2025all}, and police officers~\cite{kim2025beyond}.

Emotion work---effort to manage one's emotions and display---can be seen as one aspect of body work~\cite{debra2007what}. But body work extends further. \citet{yadav2024bodywork} illustrate this broader scope through three examples: mobile bodies in food delivery, sleep in shift work, and breast pumping in offices. \citet{shaikh2024not} examined body labor obscured by platform design, such as sanitation during COVID-19.

LPT, with its explanation of value extraction and labor transformation, is apt for directing HCI to situations where emotions and bodies are commodified, exploited, circulated, and consumed through technical means. Thinking through this implicit process of value production and realization, HCI researchers can be well positioned to devise effective interventions.

\subsection{General Approaches to Labor Studies}

HCI research on work and workplace technologies exhibits recurring patterns in engaging with labor issues. We identify three common approaches: conceptual (\textit{Surfacing Invisible Labor and Work}), empirical (\textit{Documenting Worker Experiences and Work}), and practical (\textit{Designing Interventions}). These are neither exhaustive nor mutually exclusive but illustrative of how the field engages with labor.

\subsubsection{Surfacing Invisible Labor and Work}
\label{Findings: invisible work}

Invisible work has been influential in HCI, guiding analysis of how technology systems depend on concealed human work~\cite{suchman1995making, star1999layers, nardi1999web}.
\citet{gray2019ghost}'s concept ``ghost work'' exemplifies research revealing hidden labor enabling AI. This and related work demonstrate how machine learning depends on concealed human piecework~\cite{sarkar2023enough}. This workforce enables AI functionality and profitability yet receives little from what they produce. Concealing this labor obscures its exploitation. Data annotation is just one component. HCI has investigated other hidden labor forms~\cite{boeva2023behind, catanzariti2021global}, including complex data work processes~\cite{chandhiramowuli2024making, sun2023data}, ``precision labor'' in algorithm training~\cite{zhang2025making}, red-teaming~\cite{zhang2025aura}, cloud computing~\cite{sholler2019invisible}, maintenance, and waste processing~\cite{spektor2021discarded}. Such human intervention persists because automation cannot yet handle these tasks or cannot do so cost-effectively~\cite{iantorno2022outsourcing}. \citet{fox2023patchwork} call this ``patchwork,'' the auxiliary human work required for AI to function as promised, observable in non-AI systems too~\cite{jo2025understanding, dombrowski2012labor}. Yet this labor is commodified while deskilled, marginalized, and concealed, leaving workers' precarity unaddressed.

This concealment extends beyond AI to broader precarious employment. Researchers have documented MTurk workers' additional payment management and job searching work~\cite{toxtli2021quantifying, sannon2019privacy}, gig drivers' self-tracking work to maintain accountability~\cite{hernandez2024end}, and motion capture actors' invisible contributions~\cite{cheon2024creative}. Research on invisible labor intersects with gender, disability, and migration. \citet{meissner2022hidden} examined technological work in health and beauty services. Rosner et al.~\cite{rosner2018making} recovered the invisible, women-dominated work of assembling core memory. Studies of extra work include those about disabled individuals accessing group meetings~\cite{alharbi2023accessibility, cha2024you} and disabled content creators managing hate speech and harassment~\cite{heung2024vulnerable}. \citet{zhu2025unveiling} documented immigrant care workers navigating migrant status challenges, while \citet{kojah2025dialing} examined intersectionally marginalized content creators' emotional and community work.

Beyond individual workers, invisible labor sustains human infrastructure that is visible only at breakdown~\cite{star1994steps}. When people constitute infrastructure, their labor so becomes invisible. \citet{pendse2020like} adopted this perspective in studying an Indian mental health helpline, framing volunteers' work as invisible labor. Similarly, \citet{shaikh2024not} conceptualized platform-based food delivery workers as invisible human infrastructure. \citet{akridge2024bus} examined London bus drivers' infrastructural role in ensuring passenger well-being, particularly during crises and technological breakdowns. Invisible ``workaround'' and ``restoration'' work emerge during crisis~\cite{doggett2024digital}. Sometimes this subsequent work creates new alternatives rather than restoring collapsed organizations, as seen in non-profit ambulance services mobilized by ride-hailing drivers' unions in India during the pandemic~\cite{thuppilikkat2025generative}.

Invisible labor also emerges through seemingly voluntary consumer activities on platforms. Content moderation generates economic value for platforms~\cite{ma2021advertiser, kou2017morallabor}. Additional examples include content creators' productive labor~\cite{yi2025informal, simpson2023rethinking}, publication, promotion, and articulation labor by social media artists and craftspeople~\cite{sharma2023takes, de2024poetsofinstagram, razaq2022making}, and dating app users' profile production labor~\cite{olgado2020determining}. However, many studies did not explicitly acknowledge this economic value. Unlike manufacturing or service work, these activities constitute new forms of free labor outside employment, often regarded as leisure~\cite{li2022ethical}. As capital seeks new value sources, consumers' voluntary behaviors like reviews and search histories are appropriated for profit~\cite{srnicek2017platform, zuboff2023age}. Identifying such labor—what Ekbia and Nardi call ``heteromation''~\cite{ekbia2017heteromation}—proves essential.

We distinguish invisible work from invisible labor: not all activities produce value for capital as some produce only use-value. Additionally, ``labor'' sometimes means synonymously with ``effort,'' unrelated to employment or even to work, including personal budget management~\cite{glick2022exploring}, visually impaired older adults' community participation~\cite{shinde2024designing}, Chinese citizens' coordination during COVID-19 lockdown~\cite{chen2023maintainers}, childcare workers' support-seeking on Reddit~\cite{gupta2025being}, and volunteers' maintenance of open-source software~\cite{geiger2021labor}.

Given this diverse landscape, what is the purpose of studying invisible labor? We advocate moving beyond purposes of documenting and supporting toward analyzing how capital commodifies and exploits hidden work for surplus value since profit motivates invisibilization, our analytical purpose should adjust accordingly. We return to this argument in \S\ref{Dis 5.2: valuePro}, where we propose value production as an analytical framework for invisible labor research.

\subsubsection{Documenting Worker Experiences and Work}

Much HCI scholarship documents observations, narratives, and lived experiences of workers, reflecting the field's human-centered ethos~\cite{wright2008empathy, wright2010experience} and ethnomethodological tradition~\cite{randall2021ethnography}. These studies often move from description to design recommendations without extensive analytical explanation. This approach has produced valuable knowledge about workplace technology issues, but it also leaves certain questions---particularly about causal mechanisms---less examined.

This descriptive approach often manifests as requirement generation or need-finding for system design. The studies catalog workers' motivations, experiences, and workplace problems, then derive design recommendations aimed at worker empowerment. Such studies---spanning caregiving~\cite{schurgin2021isolation}, crowd work~\cite{varanasi2022feeling}, food delivery~\cite{ma2023uncovering}, sex work~\cite{hamilton2022risk}, and beyond---offer rich accounts of worker realities, but the mechanisms producing observed conditions are seldom analyzed.

Even when studies broaden inquiry to multiple stakeholders, the pattern persists: they document each group's perceptions rather than explain why conflicts arise~\cite{martin2016turking}. For example, researchers have documented divided opinions on automation among different worker stakeholders~\cite{lee2024contrasting} and Uber drivers' frustrations whose concerns were breached by corporate actions~\cite{ma2018using}. Other studies organize findings around heterogeneous stakeholders like drivers, passengers, regulators~\cite{gloss2016designing, hsieh2023co}, but their design recommendations emphasize participant-generated solutions over analysis of why such conflicts arise. Even participatory design originally foregrounding workers' political agency and speculative design now tend to center participant narratives while leaving structural conditions less examined~\cite{cameron2021seems, ma2025speculative}.


This concern is not new to HCI. Button and Harper argued that worker narratives ``need to be used as data in the \textit{analysis} of work-practice, not as literal descriptions of work practice''~\cite[p. 267]{button1995relevance}. While their ethnomethodological alternative differs from ours, we share the diagnosis: without analytical investigation, they cautioned, derived designs risk being misled by accounts that do not reconstruct actual work practices.\footnote{Their specific concern was that user narratives and formal work documentation—e.g., order, production, and invoice procedures in manufacturing—are often post hoc reconstructions rather than faithful records of practice, warranting ethnomethodological rather than theoretical or formal investigation.} 

Analytical explication can mean explaining causes and mechanisms or interpreting meanings, as practiced in broader social sciences. LPT offers vocabulary for causal explanation. Documenting that gig workers feel ``stressed'' and use methods to cope with is descriptive; analyzing this through LPT may reveal \textit{how} platform design produces workers' consent to intensified labor despite stress, through gamified incentives, surge pricing, and metrics workers internalize as personal goals. This shifts from \textit{what} workers experience to \textit{why}, enabling different research questions: not ``how can we reduce stress?'' or ``how workers cope with stress?'' but ``how can we reveal the mechanisms producing stress?''

Some HCI work does offer mechanism analysis. Beyond the control studies discussed above (e.g.,~\cite{kristiansen2018accountability, bakewell2018everything, singh2023old, alkhatib2017examining, sehrawat2021everyday}), for instance,~\citet{muralidhar2022between} explore how food delivery workers' dependency and precarity on platforms are created, and \citet{moradi2025pseudo} provide an explanation of self-checkout kiosks' effect on retail cashiers. Kiosk design deficits reconfigure customer-cashier relations, casting customers as ``problem-makers'' and cashiers as ``police,'' leading to adversarial social interaction. Such analyses go beyond describing workplace conditions to explaining how they are produced.

\subsubsection{Designing Interventions}

Design efforts often build on empirical studies, addressing concerns about working conditions. Most focus on immediate, practical solutions, though some have informed public discourse or policy~\cite{irani2016stories}. 
Just as social work draws on sociological theory to address systemic mechanisms~\cite{wacquant2008urban}, HCI design can benefit from labor process theorization. We first survey some existing design interventions.

The first group aims to empower workers in negotiations with employers and platforms. These include tools for recording work data as evidence~\cite{khovanskaya2019tools, khovanskaya2019data}, assessing clients~\cite{irani2013turkopticon, do2024designing}, auditing algorithmic systems~\cite{calacci2022bargaining}, building data institutions~\cite{stein2023you}, and informing the public~\cite{zhang2024data}.
Other projects take less adversarial approaches: fostering worker-client relationships~\cite{bederson2011web}, accommodating diverse work styles~\cite{toxtli2024culturally}, supporting self-scheduling~\cite{uhde2021design}, or identifying design opportunities around workload and breaks~\cite{cheon2024amazon}. Other efforts include Research through Design (RtD) that provokes reflection on human labor behind IoT~\cite{desjardins2023making} and action research that cultivates awareness of how data practices harm workers~\cite{disalvo2024workers}. Designs supporting emotional labor and invisible work have also proliferated~\cite{das2025ai, chau2025all, kim2025beyond, lee2024exploring, kow2018complimenting}.

Beyond  interventions supporting individual workers, some efforts target collective action. These include tools for raising macro-level awareness~\cite{hanrahan2015turkbench}, supporting community formation~\cite{salehi2015we}, enabling local collaboration~\cite{koo2025metrics}, fostering well-being and solidarity~\cite{hsieh2025gig2gether}. Some advocate for workers' rights through co-research and democratizing algorithm design~\cite{calacci2022organizing}; others help workers document hours, exchange knowledge, and engage in collective sense-making~\cite{wolf2022designing, ming2024wage, holten2021can}. Researchers have also reflected on the ethics of positioning designers as saviors and workers as subjects~\cite{irani2016stories}.

But these efforts are often constrained by relying on participants' experiential accounts to devise solutions to systemic problems. Studies often center workers' lived experiences as the primary basis for design decisions, which can limit attention to structural mechanisms beyond participants' direct awareness. The resulting designs revolve around addressing \textit{outcomes} rather than \textit{mechanisms}. Consider wage theft: outcome-focused designs track hours and document violations—can be useful, but reactive. A mechanism-focused approach asks \textit{how} wage theft becomes possible through task fragmentation, conception-execution separation, and re-demarcation of working time. It also asks how wage theft affects work itself, for instance, increased rigidity and depersonalization in care work in accord with the wage rule~\cite{moore2017taking}. This analysis might direct design toward integrating fragmented tasks, reducing unpaid overhead time shifting between them, and enabling worker discretion over personal boundaries and work methods---addressing conditions that make wage theft possible.

Norman warned that human-centered design may not lead to the best results if designers rely solely on user inputs without analytical expertise~\cite{norman2005human}. 
LPT's core concepts give designers broader intervention space, pointing to distinct design directions. \textit{Deskilling} highlights how interfaces fragment judgment into micro-tasks, suggesting designs that preserve skill integration. \textit{Separation of conception from execution} highlights how workers lose control over planning, opening possibilities for tools returning autonomy over planning and method to workers. \textit{Consent manufacturing} reveals how workers embrace intensified labor as self-achievement, prompting designs that uncover managerial origins of gamified incentives. These represent different \textit{levels} of intervention, from task design to process control to ideological conditions of work.

Even interventions that fall short of their goals yield reflection on why certain approaches did not work as intended, expose limitations in the theoretical assumptions guiding design, and inform the community. HCI researchers can use design not only as intervention but as provocation, making hidden norms visible, as for breaching experiments~\cite{crabtree2004design}. Engagement with the labor process facilitates breaching experiments that alter mechanism factors—then analyze differences—to reveal the often taken-for-granted workplace norms and labor arrangements.

Just as ethnography sensitizes designers to context~\cite{randall2021ethnography, blomberg2013reflections}, LPT can cultivate labor sensitivities. Many designs optimize efficiency and managerial control at workers' expense. Understanding mechanisms producing exploitative or harmful working conditions may reduce designers' unconscious reinforcement of managerial intent. LPT also reveals structural constraints, such as profit imperatives and managerial resistance, on implementing human-centered technologies to improve working conditions, encouraging more realistic design expectations.


\section{Future Agenda for Research on Work and Labor} 
\label{Discussion}

The review of HCI scholarship indicates patterns and trends of studies around labor and technology. There are merits of HCI's approaches to labor and work (integration of sociotechnical factors and design), but also limits in need of greater attention. We now turn to how LPT poses a provocative agenda for HCI. While HCI offers diverse accounts of worker experiences, LPT foregrounds how labor is contained within regimes of control and extraction. This perspective suggests ways in which HCI might deepen engagement with the structures of work, broadening analytical scope and opening space for theory-informed design. We begin by disentangling two basic concepts: labor and work.

\subsection{Distinguish Labor From Work}

A central motivation for ``back to labor'' is revitalizing political economic analysis of work under capitalism~\cite{greenbaum1996back}. This requires clarifying key terms---especially distinguishing \textit{labor} from \textit{work}. In HCI, labor and work are often used interchangeably, though Greenbaum warned that ``labor is more than work''~\cite[p. 229]{greenbaum1996back}. While both denote human effort, \textit{labor} carries political economic meaning—connoting not just activity but its social relations of production.

\textit{Labor} is foundational to political economy. For neoclassical economists, labor equals work: tasks performed for a ``fair'' wage, with value created through market exchange~\cite{foley1986understanding}. Marx developed the labor theory of value as a systematic critique of capitalism, distinguishing his position from (neo)classical economists. Marx argued that market exchange creates no new value at the societal level—value is created only through human labor, and capitalists accumulate capital by extracting surplus in production. Surplus extraction is the crucial inference from Marx's theory. From this view, \textit{labor} is the expenditure of labor power producing value appropriated for accumulation.\footnote{The labor here is only productive labor. Unproductive labor does not produce value or commodities for capital accumulation, but may only produce use-value for capital, self-consumption, or other non-market purposes~\cite{foley1986understanding}.} \textit{Work}, by contrast, lacks this framing and can obscure the unequal exchange embedded in wages.

\begin{figure}[htbp]
    \centering
    \includegraphics[width=0.8\columnwidth]{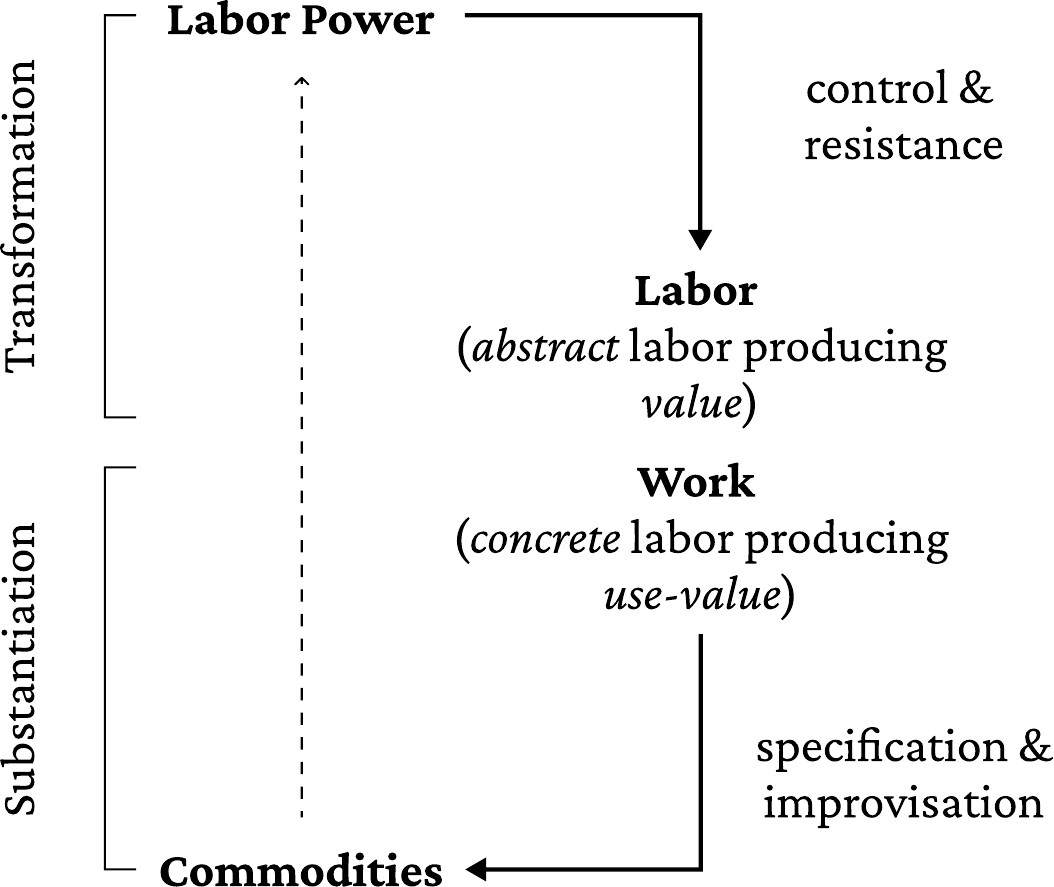}
    \caption{Analytical distinction between labor and work. Both refer to human effort in production, but emphasize different facets. In Marxist terms, labor captures the abstract, \textit{value}-producing aspect of human effort, while work refers to its concrete, \textit{use-value}-producing aspect. The transformation process (Labor Power → Labor) foregrounds managerial control and worker resistance—the domain of LPT. The substantiation process (Work → Commodities) foregrounds task specification and worker improvisation—the domain of conventional work practice studies. The dashed arrow indicates how efficient commodity production reduces the cost of reproducing labor power, intensifying exploitation (see \S\ref{S2.1: hiddenAbode}).}
    
    \Description{The diagram shows the analytical distinction between labor and work. Labor is abstract labor producing value while work is concrete labor producing use-value. They emphasize two parallel facets in the production process---one of labor power transforming into labor, contested by managerial control and labor resistance. This corresponds to an arrow on the diagram pointing from labor power to labor. The other facet is about work substantiating commodities through product specification and worker improvisation. This corresponds to an arrow on the diagram pointing from work to commodities. More efficient commodity production decreases the cost and value of reproducing labor power at the societal level, which enables more exhaustive use of labor power. This corresponds to a dashed arrow pointing from commodities to labor power.}
    \label{laborPower}
\end{figure}

Shifting frameworks changes what becomes salient. As HCI scholars have noted~\cite{greenbaum1996back, tang2023back, gloss2016designing}, focusing on work practice---what tasks are and how they are performed---tends to sideline wage-labor relations. HCI studies of programmers often characterize work practices and tool use, then seek improvement through design guidelines or prototypes. The goal of such studies is to streamline work and raise productivity---creating more products from a given workload---a focus already distant from wage-labor relations. This focus corresponds to the ``substantiation'' process in \autoref{laborPower}. 
LPT takes a different approach, treating programming as wage labor~\cite{wu2024play}. Instead of optimizing task performance, it explains how relations among workers, managers, and external actors---platforms, clients, regulators---structure working conditions~\cite{greenbaum1996back}. This distinct attention corresponds to the ``transformation'' process in \autoref{laborPower}. 

The distinction between labor and labor power further clarifies the analytical focus of labor-oriented studies.
Technological advancement affects labor and labor power differently. HCI designs, as potential commodities, can devalue and exhaust labor power even while enhancing labor's productive force (see \autoref{laborPower}). This paradox exposes what uncritical analysis obscures: exploitation. This recognition establishes a moral obligation: knowledge and design should respond to labor power conditions, not just labor productivity. Consider a tool designed to optimize crowd worker selection and management. Evaluated by labor productivity, it might succeed---tasks assigned, information exchanged, coordination seamless. But viewed through LPT, labor power is confined to homogeneous micro-work, rendered disposable, insulated from conception and the final product. Fortunately, we can always imagine automation differently: coordination tools that ensure peer production over hierarchy, broad worker participation over task segregation.

Labor and work mark two divergent paths for HCI, leading to distinct knowledge production. Attention to labor power reminds us how technologies are situated within the dual process of enhancing productivity and intensifying exploitation.

\subsection{Link Work Practice to Value Production}
\label{Dis 5.2: valuePro}

HCI's focus on work practice frames work as producing \textit{use-value}. This may explain why the field struggles to engage deeply with labor~\cite{gloss2016designing}, a concept more concerned with \textit{value}. With few exceptions (e.g.,~\cite{li2022ethical}), HCI has paid limited attention to how work produces economic value. This is where LPT becomes relevant. Whereas use-value production takes countless forms depending on the commodity (e.g., baking bread, writing code, cleaning offices), the production and extraction of value follows a more general logic under capitalism---surplus value appropriated as capital. HCI scholarship could benefit from linking work practice accounts to political economic analysis of value.

The proliferation of studies on ``invisible'' labor reflects growing awareness of overlooked work, especially as sociotechnical systems become more complex and opaque. Invisible work refers to invisibility of the work, the worker, or both~\cite{star1999layers, nardi1999web, poster2016introduction}. Much HCI scholarship focuses on the former (see \S\ref{Findings: invisible work}). Scholars have offered two main rationales for such studies: modeling work practice for better design fit, and recognizing devaluation to combat injustice~\cite{star1999layers}. LPT provides another rationale from political economy: revealing how capital \textit{commodifies} invisible work so that we can mitigate exploitation. 
This reframes positive and moral questions as political economic ones—asking not just what is hidden, but how and why it becomes a site of extraction.
Consider emotional labor. Flight attendants' smile and bill collectors' urgency toward debtors are examples of emotion work that has been commodified because they attract consumption or generate profit, thereby becoming emotional labor~\cite{hochschild2012managed, sutton1991maintaining}. Commodification is a necessary step for surplus value to be extracted from labor. This raises the question of whether streamlined workflows and public recognition adequately address what workers need—or whether improved wages and working conditions deserve greater attention. Taking labor seriously requires connecting work practice to commodification and surplus value extraction. Only then can concepts like ``invisible labor'' inform interventions in value chains.

\subsection{Study Up the Management}
\label{Dis 5.3: management}

We recommend that HCI researchers examine the management and control of labor holistically. A workplace regime, defined as ``systematic patterns of managerial control''~\cite[p. 1]{wood2021workplace}, encompasses not only mechanisms such as monitoring and discipline but also social, administrative, and reproductive control forms. While prior HCI work has analyzed algorithmic and bureaucratic control~\cite{anjali2021watched}, subtler control forms receive less attention~\cite{singh2023old}. Social control includes peer pressure, reputational metrics, and mutual surveillance that discipline conduct under the appearance of autonomy~\cite{greenbaum1988search}. Administrative control encompasses the ``auxiliary’’ workflows of check-ins, leave requests and sanctions, training, HR management, and accounting, which are increasingly automated and embedded in everyday coordination~\cite{chen2024digitalrun}. Reproductive control extends managerial reach into the sphere of labor power reproduction—housing, meals, rest—as in dormitory labor regimes in Chinese factories, where workers live under employer oversight~\cite{smith2006dormitory}. Revealing these less visible mechanisms brings us closer to holistic accounts of workplace regimes.

Understanding these control forms requires examining their underlying logic. LPT views management as capital's imperative to extract labor from indeterminate labor power—and labor power as continually resisting~\cite{edwards1979contested}. Recognizing this tension as the driving force behind workplace technologies enables HCI researchers to ground critique in political economy and identify intervention opportunities---a return to materialist analysis of workplace struggle. While much HCI work centers ethical concerns about workplace technologies (e.g., privacy and surveillance~\cite{kapoor2022weaving}), attention to workers' material struggle against extractive labor arrangements is no less important~\cite{ekbia2015political}. This stands apart from conventional management studies that frame labor as a resource to be ``optimized'' for profit.

We also call for studying management as labor. Middle-layer managers—those occupying supervisory roles yet employed as wage laborers—often perform what Marx called ``unproductive labor''~\cite[p. 152]{marx1969theories}. They do not generate surplus value \textit{directly} but produce use-value—reporting, allocation, mediation—with utility for business operations and accumulation. Despite their position, middle managers can be underpaid: their labor time produces more value than they receive in wages~\cite{mccann2008normalized, foley1986understanding}. Their ambivalent role—between capital and workers—makes them strategic for investigation. Studying their practices, tools, and labor relations opens new lines of inquiry for HCI. Rethinking control holistically and attending to management as labor enriches our understanding of work, expands intervention scope, and aligns HCI with critical labor studies.

\subsection{Analyze Consent and Legitimacy}

Much HCI scholarship on control focuses on coercion—enforcement through surveillance, evaluation, and discipline; workers are cast as ``resistive subjects''~\cite[p. 167]{hall2010renewing}. These approaches, grounded in first-wave LPT, effectively expose how management undermines rights and autonomy. But this framing remains limited. Coercion is bounded by worker endurance and acceptability. Consent often supplements coercion to extend beyond these limits. As Burawoy notes~\cite{burawoy1990politics}, coercion and consent are not mutually exclusive: even the most coercive systems require worker cooperation. At the same time, techniques appearing to promote autonomy often originate in top-down control. Many contemporary controls—especially algorithmic—operate through subtle forms of inducement, participation, and alignment in addition to enforcement. 
Second-wave theorists, including Burawoy, emphasize how consent reproduces labor relations—not through false consciousness or deception but through negotiations, routines, and performances aligning worker interests with organizational objectives.

Yet despite its relevance, consent remains under-theorized in HCI, with few exceptions~\cite{singh2023old, cheon2025fulfillment}. Even when implicit controls like gamification are examined, they are rarely framed as consent mechanisms~\cite{ramesh2023ludification}. A persistent control/resistance binary reinforces this, presuming rigid opposition between worker and managerial interests. This obscures how worker participation and autonomy can be manufactured by control strategies. As labor scholar Edwards~\cite{edwards1990understanding} urges, the boundaries of control and resistance should expand. We extend this call to HCI. This shift—from coercion to consent, despotism to hegemony—does not dismiss agency but interrogates how agency is cultivated and governed. Beyond documenting worker experiences of surveillance and discipline, we should examine the sociotechnical conditions under which workers participate in their own subordination—how control is legitimized~\cite{burawoy1979manufacturing}. HCI's empathic strength can expose these dynamics, but only by engaging analytically with how worker participation can be shaped---or co-opted---by managerial strategies. Attending to how consent is organized and sustained can illuminate subtle control and identify sites of intervention. By attending to how consent is legitimated, organized, and sustained, we can illuminate more subtle control mechanisms—and identify concrete sites for intervention.

\subsection{Move Beyond the Point of Production}

The above  LPT-informed suggestions have been anchored primarily at the point of production---a workplace regime that was central to early LPT but is increasingly limited under globalized capitalism ~\cite{thompson2004labor}. As capital expands through outsourcing and offshoring of labor power and supply chains—facilitated by digital technologies—value production is no longer confined to a single worksite; it unfolds across distributed global value chains. Viewing workplace technologies within global production therefore expands intervention sites beyond task-level redesign to standards, contracting terms, and cross-organizational governance. These distributed arrangements call for broader attentions. HCI scholarship should attend to labor in sociotechnical systems operating across regional, national, and transnational circuits of value production and extraction. While crowd work and freelancing have been the primary focus of HCI research in this regard~\cite{martin2016turking}, similar transformations in labor organization are reshaping manufacturing, service work, and domestic work; logistics and labor mobility can reveal how exploitation operates across distributed sites~\cite{moore2018paying, smith2010go}.

Beyond spatial expansion, market forces also matter. LPT has been criticized for privileging production over other circuits—such as labor and commodity markets—that shape workplace conditions~\cite{kelly1985management}. Abundant labor supply depresses wages, and fluctuations in product demand may encourage flexible forms of production organization~\cite{vallas1999rethinking}. The circuit of production is not isolated. For HCI, focusing only on immediate worksites risks attributing workplace problems to local tools and interfaces when system-level constraints are set by labor and commodity markets (e.g., fluctuating wages driven by global competition) or by upstream procurement and platform policies (e.g., minimum contract volume requirements). We therefore call on the HCI community to attend to technologies operating beyond production—particularly in the external labor market (e.g., job-seeking platforms and recruitment systems~\cite{cherubini2021elucidating, sharma2023takes, lu2021uncovering}), internal labor market (e.g., systems for training, promotion, and layoffs~\cite{littler1990labour}), and commodity markets (e.g., targeted marketing infrastructures). Understanding how technologies mediate these broader economic processes helps explain how they shape workers' lived experiences. Furthermore, as Burawoy's comparative analysis of factory regimes in \textit{The Politics of Production }~\cite{burawoy1990politics} suggests, external factors such as enterprise-state relations, a nation's position in the global capitalist system~\cite{immanuel1974world}, and the social reproduction of labor power (housing, care, daily sustenance) also structure the specific forms of conflict at the production site~\cite{burawoy1990politics}. We thus encourage HCI researchers to situate workplace technologies within broader circuits of production, exchange, and reproduction to generate more situated and actionable insights.

\subsection{Design Alternative Institutions}

HCI design researchers \citet{gloss2016designing} note the difficulty of directly linking design with labor issues. Rather than sidestepping this difficulty, we pose a question for HCI: What would it mean to design for labor under---and, where necessary, against---the constraints of capitalism?

Addressing this challenge, \citet{wolf2022designing} proposed a dual tactic–strategic approach: designing for immediate concerns (e.g., wage theft and unsafe working conditions) while also imagining long-term alternatives that enable new institutions and futures. We echo their call but urge HCI researchers to experiment with designs that mitigate value extraction, deskilling, fragmentation, and alienation---that LPT has analyzed as endemic to capitalist labor processes. Grounding such designs in LPT helps identify where to intervene---targeting the specific mechanisms (e.g., deskilling, ownership concentration, conception-execution separation) through which exploitation operates. For instance, LPT highlights how platform owners' exclusive ownership of infrastructural resources can turn workers into underpaid or unpaid laborers---an insight that points toward shared platform ownership, co-governance, and fair value distribution as design opportunities.

Such efforts are already emerging. Platform cooperatives have gained traction, yet have received limited attention in HCI~\cite{lampinen2018member, beloturkina2025charting}. Related proposals include cooperative AI models collectively owned by crowd workers~\cite{sriraman2017worker}, data cooperatives~\cite{solano2025running}, and initiatives ``commoning'' goods and services~\cite{chauhan2024commoning, sciannamblo2021caring}. Despite market challenges~\cite{pencavel2002worker}, these approaches emphasize shared ownership and co-governance, offering alternatives to nominally ``democratic'' participation arrangements in which workers may be co-opted~\cite{ahmed2022owns, harrington2019deconstructing}. We call on the HCI community to engage with these alternatives—not only as one-off projects, but as starting points for institutional change. 

\subsection{Unnaturalize Bourgeois Designs}

As early as 1982, a CHI paper warned that human factors research mirrored scientific management—both aiming to reduce human unpredictability and eventually eliminate labor~\cite{kraft1982human}. Yet this history seems forgotten as the article remains uncited as of this writing. Management-driven designs continue deepening labor fragmentation, alienation, and precarization~\cite{valentine2017flash, musthag2013labor, dow2012shepherding, harris2015effects}. Mainstream designs align with ``neoliberal capitalist visions of a hi-tech future''~\cite{feltwell2018grand} and technosolutionism prioritizing efficiency and acceleration~\cite{lin2021techniques}. HCI knowledge sometimes comes from perspectives that take capitalism and existing labor relations for granted, treating them as natural or unquestionable. These tendencies call for unnaturalizing bourgeois designs---questioning the class interests embedded in design assumptions and exposing what is presented as inevitable as historically contingent.

This unnaturalizing work requires reshaping the political economy of HCI design and knowledge production. This is no easy task given HCI as an institution is entangled with capitalism---through industrial funding, vendor partnerships, and evaluation metrics oriented toward productivity and market success. But change can begin with research practice.
HCI researchers must realize human-centered design is a double-edged sword. Productivity gains may heighten worker competition; communication tools may enable workplace surveillance; flexibility may produce employment precarity; bureaucratic pipelines may reinforce worker compliance. HCI design, as well as knowledge, is always situated in particular contexts and perspectives. Researchers must ask: Will designs turn against workers? Whose interests do they serve?  Researchers' own positionality---their relationship to capital, labor, and the communities they study---should be stated and contested whenever possible; sensitivity to how designs affect labor conditions should be cultivated. However natural they may appear, labor relations under capitalism are organized through surplus extraction---and this affects us all.
\section{Conclusion}

2024 marks the 50th anniversary of \textit{Labor and Monopoly Capital}’s publication. Over the past five decades, the world economy---and with it, work organization---has undergone profound transformations, including post-Fordism, lean production, mass service work, knowledge economy, global labor arbitrage, and platformization. Throughout these shifts, Labor Process Theory has remained vital, evolving as a critical tradition for analyzing work’s changing dynamics under capitalism. 
At the heart of labor process analysis is machinery, reflecting capitalism's enduring drive to revolutionize production. These technological developments confront workers as a persistent material reality. Yet technology is not inherently deterministic or detrimental, especially when design decisions are made with collective critical consciousness of labor conditions. In this sense, HCI scholarship is well-positioned to engage labor process issues—not only tracing their shifting contours, but co-constructing political economic futures of work through critique, debate, and design.

\begin{acks}
The authors are grateful to the anonymous reviewers who have offered valuable comments and critiques to this paper. Thanks to Pyeonghwa Kim for her insights into the literature and to others who have read drafts of the paper.
\end{acks}

\bibliographystyle{ACM-Reference-Format}
\bibliography{main}

\end{document}